\documentclass[useAMS,usenatbib]{mn2e}

\usepackage{graphicx} \usepackage{epsf} \usepackage{lscape}
\usepackage{lipsum, enumitem}
\usepackage{amsmath} 
\newcommand{\angstrom}{\text{\normalfont\AA}}
\newcommand{\F}{$\mathcal{F}$}
\newcommand{\dF}{$\Delta\mathcal{F}_{\rm Ca}$}

\title[Sounding stellar cycles with {\it Kepler} -- II]{Sounding stellar cycles with {\it Kepler} -- II. Ground-based observations\footnote{Based on observations made with the Nordic Optical Telescope, 
operated on the island of La Palma jointly by Denmark, Finland, Iceland, 
Norway, and Sweden, in the Spanish Observatorio del Roque de los 
Muchachos of the Instituto de Astrof{\'i}sica de Canarias.}}
\author[C. Karoff et al.]{C.~Karoff$^{1}$\thanks{E-mail: karoff@phys.au.dk}, 
T.~S.~Metcalfe$^{1,2}$,
W.~J.~Chaplin$^{3}$,
S.~Frandsen$^{1}$,
F.~Grundahl$^{1}$, \and H.~Kjeldsen$^{1}$,
J.~Christensen-Dalsgaard$^{1}$,
M.~B.~Nielsen$^{1,4}$,
S.~Frimann$^{1}$, \and A.~O.~Thygesen$^{5}$,
T.~Arentoft$^{1}$,
T.~M.~Amby$^{1}$,
S.~G.~Sousa$^{6}$,
D.~L.~Buzasi$^{7}$\\
$^{1}$Stellar Astrophysics Centre, Department of Physics and Astronomy, Aarhus University, Ny Munkegade 120, DK-8000 Aarhus C, Denmark\\
$^{2}$Space Science Institute, 4750 Walnut Street, Suite 205, Boulder, Colorado 80301 USA\\
$^{3}$School of Physics and Astronomy, University of Birmingham, Edgbaston, Birmingham, B15 2TT, UK\\
$^{4}$Institut F\"{u}r Astrophysik, Georg-August-Universit\"{a}t, Friedrich-Hund-Platz 1, 37077 G\"{o}ttingen, Germany\\
$^{5}$Centrum f\"{u}r Astronomie der Universit\"{a}t Heidelberg, Landessternwarte, K\"{o}nigstuhl 12, 69117 Heidelberg, Germany\\
$^{6}$Centro de Astrof{\'i}sica, Universidade do Porto, Rua das Estrelas, 4150-762, Porto, Portugal\\ 
$^{7}$College of Arts and Sciences, Florida Gulf Coast University, 10501 FGCU Boulevard South, Fort Myers, FL 33965-6565, USA\\
}
\begin{document}

\date{Accepted ??. Received ??}

\pagerange{\pageref{firstpage}--\pageref{lastpage}} \pubyear{2012}

\maketitle

\label{firstpage}

\begin{abstract}
We have monitored 20 Sun-like stars in the $Kepler$ field-of-view for excess flux with the FIES spectrograph on the Nordic Optical Telescope since the launch of $Kepler$ spacecraft in 2009. These 20 stars were selected based on their asteroseismic properties to sample the parameter space (effective temperature, surface gravity, activity level etc.) around the Sun. 
Though the ultimate goal is to improve stellar dynamo models, we focus the present paper on the combination of space-based and ground-based observations can be used to test the age-rotation-activity relations.  

In this paper we describe the considerations behind the selection of these 20 Sun-like stars and present an initial asteroseismic analysis, 
which includes stellar age estimates. We also describe the observations from the Nordic Optical Telescope and present mean 
values of measured excess fluxes. These measurements are combined with estimates of the rotation periods obtained from a 
simple analysis of the modulation in photometric observations from $Kepler$ caused by starspots, and asteroseismic determinations of stellar ages, to test relations between between age, rotation and activity. 
\end{abstract}


\section{Introduction}
Some of the greatest advances in our understanding of the solar dynamo 
during the last few decades have been brought about with the aid of 
helioseismology. In particular, the mapping of differential rotation 
inside the Sun \citep{1998ApJ...505..390S} and constraints on meridional 
circulation \citep{1996Sci...272.1306H} have helped push forward 
this understanding. Unfortunately, our inability to make reliable predictions of 
the evolution of the solar cycle in the transition between solar cycles 23 and 24 implies that solar 
dynamo models have still not reached a stage where they can be used for 
predicting the solar cycle.

Observations of activity cycles in Sun-like stars present an excellent 
opportunity to improve our understanding of the solar dynamo 
\citep[see e.g.][]{2008ssma.book.....S}. Asteroseismic observations of 
activity cycles in Sun-like stars can facilitate this understanding 
because they allow us to compare the changes taking place 
in the interior of the stars to the changes taking place on the 
surface, as discussed by \citet[][hereafter CK09]{2009MNRAS.399..914K}.

Most of the known activity cycles in Sun-like stars were detected first 
from Mount Wilson Observatory \citep{1978ApJ...226..379W, 
1985ARA&A..23..379B, 1995ApJ...438..269B} and later from Lowell 
Observatory \citep{2007AJ....133..862H}. These detections have revealed 
that around half of the observed Sun-like stars show clear periodic cycles, 
with periods between 2.5 and 25 years \citep{1995ApJ...438..269B}. 
\citet{1998ApJ...498L..51B} and \citet{1999ApJ...524..295S} showed how 25 stars with well-defined periods can 
be separated into two groups: a group of young active stars, and a group of older inactive stars. The stars in 
the former group have cycle periods that are typically 300 times longer than their rotation periods, while the 
stars in the latter group have cycle periods that are typically only 100 times longer. \citet{1981PASP...93..537D} (see also
\citet{2007ApJ...657..486B, 2008LRSP....5....2H}) suggested that 
this bifurcation, which is known as the Vaughan-Preston gap 
\citep{1980PASP...92..385V}, is a consequence of the dynamo being seated at 
two different places inside young and old stars.

The essential hypothesis of \citet{1981PASP...93..537D} and \citet{2007ApJ...657..486B} is that Sun-like stars will arrive at the main sequence with 
a nearly homogenous distribution of interior angular momentum.
This means that the largest change in the radial rotation rate -- the 
strongest radial shear layer -- is found just below the surface of these 
stars. As the stars evolve they lose angular momentum to stellar 
winds and if it is assumed 
that the loss of angular momentum from the surface of these stars only 
affects the outer convection zone, it follows that a strong shear layer 
will develop near the bottom of the convection zone -- creating a 
so-called tachocline. This is of course a simplified description of the evolution of angular momentum in Sun-like stars. A more detailed description which includes core-envelope decoupling and disk interactions can be found in e.g. \citet{2001ApJ...561.1095B, 2003ApJ...586..464B, 2007ApJ...669.1167B, 2010ApJ...722..222B}.

The evolution of stellar angular momentum also has
implications for stellar activity levels, as first suggested by \citet{1972ApJ...171..565S}. These 
relations imply that stars not only lose angular momentum to stellar 
winds as they grow older, but they also become less active as they 
spin down. \citet{1972ApJ...171..565S} based his suggestion on 
observations of the Sun and of three stellar clusters (the Pleiades, Ursa 
Major and the Hyades) for which ages could be estimated at the time. Since then large efforts have been invested in improving these relations by measuring rotation periods in stellar clusters with known ages \citep[e.g.][]{1982Msngr..28...15V, 1985ApJ...289..247S, 1993ApJ...402L...5S, 1996A&A...305..498A, 1999ApJ...516..263B, 2001ApJ...563..334S, 2002ApJ...576..950T, 2009ApJ...691..342H, 2011ApJ...733L...9M, 2011ApJ...733..115M}. 

Asteroseismology offers a unique tool to address this problem because it allows us to measure reliable ages of field stars, independent of rotation period and activity level. If we can also measure the rotation period and activity levels of these stars, they can be used to test and improve the age-rotation-activity relations. In this study, we present age-rotation-activity relations for 10 Sun-like stars with ages between one and 11 Gyr based on asteroseismic measurements.

The shortest activity cycle period measured from the programmes at the Mount 
Wilson and Lowell observatories is 2.52 years, but three recent results 
have revealed that Sun-like stars can have even shorter ($<$ 2 years) cycle periods as well.  \citet{2010Sci...329.1032G} detected the 
signature of a short magnetic activity cycle (period between 120 days 
and 1 year) in the F5V star HD~49933 using asteroseismic measurements 
from the CoRoT satellite \citep{2008A&A...488..705A}. \citet{2010ApJ...723L.213M} discovered a 1.6 year 
magnetic activity cycle in the F8V exoplanet host star $\iota$ Horologii 
using synoptic Ca HK measurements obtained with the Small and Moderate 
Aperture Research Telescope System 1.5 m telescope at Cerro Tololo 
Interamerican Observatory and concluded that if short magnetic activity 
cycles are common, NASA's {\it Kepler} mission should detect them in the 
asteroseismic measurements of many additional stars. It is also worth noting that
\citet{2010ApJ...718L..19F} found evidence of a quasi-biennial solar cycle in 
the residuals of oscillation frequency shifts measured by the Birmingham 
Solar-Oscillations Network \citep[BiSON,][]{1995A&AS..113..379E} and by the 
Global Oscillations at Low Frequencies 
\citep[GOLF,][]{1995SoPh..162...61G} instrument on board the ESA/NASA 
Solar and Heliospheric Observatory (SOHO) spacecraft. They speculated that 
this quasi-biennial cycle might be driven by the near surface shear 
layer -- in contrast to the 11-year cycle which is believed to be driven by 
the tachocline. 

These examples of stars with short ($<$ 2 years) cycle periods show that stars can have short cycles, but two stars is too few to reliably judge whether or not short cycle periods are common in Sun-like stars. On the other hand, the results on HD 49933 and $\iota$ Horologii do indicate that dynamos may be fundamentally 
different in F stars compared to the Sun and other G and K 
main-sequence stars, most likely due to the thin outer 
convection zones in F stars. It should be noted here that F-type stars are under-represented in both the Mount Wilson and Lowell samples as the focus of these surveys was on Sun-like stars -- in fact no stars earlier than F5 were on the original Mount Wilson list \citep{1978ApJ...226..379W} and this might explain why short periodic cycles were not seen by the Mount Wilson and Lowell surveys.

The layout of the rest of the paper is as follows. In Section~2 we 
describe the programme, including the target selection and the 
observations that have been conducted so far. An analysis of these 
observations can be found in Section~3, and results on both the activity 
distributions and the age-rotation-activity relations are provided in 
Section~4. Section~5 includes a discussion of these results.

\section{programme description}
The programme {\it Sounding Stellar Cycles With Kepler} combines 
high-precision photometric observations from {\it Kepler} with
ground-based spectroscopic observations from the Nordic Optical 
Telescope (NOT). The number of targets in the programme was 
based on the need to have enough stars to cover both sides of the
Vaughan-Preston gap \citep{1980PASP...92..385V} and to adequately sample 
the rotation period vs. cycle period diagram by B{\"o}hm-Vitense. These
requirements were tempered by the need to obtain the necessary
observing time at 
the NOT each summer and the desirability of observing these targets for the entire 
length of the {\it Kepler} mission, which includes both the nominal and 
the extended missions (CK09). Based on these considerations, we decided that the 
programme should include 20 targets.
 
Ideally, the targets should have been selected prior to the launch 
of $\it Kepler$ so that observations at 
the NOT could begin at the same time. We initially tried to 
follow this approach by selecting targets based on their magnitudes 
and colours in the Kepler Input Catalog 
\citep[KIC,][]{2011AJ....142..112B}, but when we received the first 
observations from {\it Kepler} it was clear that the targets selected 
prior to launch were not ideal -- i.e. the stars did not show oscillations. There are likely two reasons for this: firstly,
the stellar properties measured with asteroseismology turned out to be 
different \citep{2011ApJ...738L..28V} from the less precise parameters implied by the KIC, and secondly, our predictions of how activity and other factors affected the asteroseismic signals were not good enough prior to the launch of {\it Kepler} 
\citep{2011ApJ...732L...5C}. We therefore 
selected a new set of targets in 2010 using the first asteroseismic results from 
{\it Kepler} \citep{2011Sci...332..213C}. 

\subsection{Target selection}
The 20 stars in the programme are shown in Figs.~1~\&~2 along with the same 
Padova isochrones \citep{2004A&A...415..571B, 2002A&A...391..195G, 
2004A&A...422..205G} that were used in CK09 -- the isochrones were 
calculated for 6 different ages (1.0, 1.6, 2.5, 4.0, 6.3 \& 10.0 Gyr), using a metallicity of $Z$ = 0.02. Though the structure and location of the isochrones does depend on the metallicity \citep{2008A&A...484..815B}, it seems safe to conclude that none of the stars have evolved far beyond the main sequence. The names, positions and 
magnitudes of the 20 stars are given in Table~1 and stellar parameters are found in Table~2.

The basic principles guiding our selection process were: 

\begin{enumerate}
\item[]{(i)} Use stars brighter than 10th magnitude observed in the first 3 months of the {\it Kepler} survey phase as candidates.
\item[]{(ii)} Preferentially select cooler stars.
\item[]{(iii)} Ensure that oscillations can be seen in the acoustic spectrum.
\item[]{(iv)} Ensure that the oscillation modes can be understood in the framework of the asymptotic frequency relation.
\item[]{(v)} Ensure that hints of rotational splitting can be seen. 
\item[]{(vi)} Ensure that the small separation is relatively large (6$\mu$Hz).
\item[]{(vii)} Ensure that both active and inactive stars have been selected (based on ground-based observations of chromospheric activity in the  stars).
\end{enumerate}

The only significant difference between the basic principles given in CK09 and the ones actually used was that, due to the fact that {\it Kepler} has been 
able to do much better photometry on saturated stars than expected 
\citep{2010ApJ...713L.160G}, stars as bright as 6.9 were also 
selected (as can be seen in Table~1). In order to evaluate the 
small frequency separation and thus the ages of the stars independently from prior investigations, we calculated the autocovariance of each power spectrum \citep{2009A&A...506..435R, 
2010MNRAS.408..542C, 2010arXiv1003.4167K}. 

When we formulated the basic principles for selecting targets, we expected that the stars would be observed for three months
in the survey phase of the mission. However, this was changed to only one month \citep{2010AN....331..972K}. We therefore did not require that rotational splitting could be seen in the spectra calculated from only one month of observations. Instead, we ensured that stars showing rotational modulation from spots in their light curves were selected along with stars that did not show any modulation.

\subsection{Asteroseismic results}
The calculated autocovariances of the spectra were only used for 
calculating the small frequency separations and thus to guide the target 
selection \citep{2009A&A...506..435R, 
2010MNRAS.408..542C, 2010arXiv1003.4167K}. The stellar properties were inferred with the SEEK package \citep{2010ApJ...725.2176Q}. The SEEK package uses a large grid of stellar models computed with the Aarhus STellar Evolution Code \citep[ASTEC;][]{2008Ap&SS.316...13C}. To identify the best model, SEEK compares the observational constraints (large and small frequency separations, effective temperature and metallicity) with every model in the grid and makes a Bayesian assessment of the uncertainties. The average large and small frequency separations were obtained as simple mean values of the individual oscillation frequencies from \citet{2012A&A...543A..54A}; these frequencies are measured using 9 months of observations  -- March 22, 2010 to December 22, 2010. The effective temperatures and the metallicities were obtained from \citet{2012MNRAS.423..122B}.

For two stars, we were unable to proceed with the asteroseismic analysis in the standard manner.
KIC 10124866 turned out to be an asteroseismic binary with two sets of 
oscillation frequencies in the acoustic spectrum, which complicates 
the analysis. A dedicated paper is therefore in 
preparation by White et al. for this star and no results are thus 
presented here.

KIC 4914923 was not among the 61 stars analysed by 
\citet{2012A&A...543A..54A}, but it was analysed in the same way as part of 
more recent work by the same group.
The asteroseismic results are presented in Table~2.

\subsection{Observations}
The photometric asteroseismic observations are described by 
\citet{2012A&A...543A..54A}.

The ground-based observations were obtained with the high-resolution 
FIbre-fed Echelle Spectrograph (FIES) mounted on the 2.6 meter Nordic 
Optical Telescope \citep{2000mons.proc..163F}. Sufficient time was 
awarded to obtain spectra on three epochs for each star during each
year of the nominal $Kepler$ mission in 2010, 
2011 and 2012. Observations were obtained using the low-resolution 
fiber (R=25000). The epochs were typically placed in April, June and 
August. The spectra were obtained with 7 minute exposures resulting in 
a S/N above 100 at the blue end of the spectrum for the faintest stars. A 
few stars are missing observations at one or more epochs -- either due to bad 
weather or passing clouds or due to cosmic ray hits, but 
most stars have observations at most epochs.

As described above we used effective temperatures and the metallicities from \citet{2012MNRAS.423..122B} and oscillation frequencies from \citet{2012A&A...543A..54A} in the asteroseismic analysis.  For the analysis of the spectra we used $B-V$ measurements of the stars from \citet{2000A&A...355L..27H}, which are listed in Table~1, together with Hipparcos luminosities for the Hipparcos stars in the sample.

\subsection{Data reduction}
The reduction of the spectra, which includes bias and flat-field 
subtraction, blaze correction, wavelength calibration and removal of 
cosmic ray hits, was done using 
FIEStool\footnote{http://www.not.iac.es/instruments/fies/fiestool/FIEStool.html}. 
FIEStool returns 1-D echelle spectra. We then merged the orders that 
covers the range between 3885 and 4015 {\AA} and cross-correlate these 
merged spectra with a solar spectrum to place the observed spectra on 
a reference wavelength grid with velocities zeroed.

\section{Analysis}
In order to calculate the excess flux from the stars \dF (defined as the surface flux arising from magnetic sources) we have followed 
\citet{2007AJ....133..862H} as closely as possible. This recipe contains 
the following steps (described in more detail below):

\begin{enumerate}
 \item[] {(i)} Correct for blanketing
 \item[] {(ii)} Normalize the spectra to an absolute flux scale
 \item[] {(iii)} Measure the flux in a 1{\AA} bandpass $\mathcal{F}_{\textup{1\angstrom}}$ 
 \item[] {(iv)} Correct for photospheric flux $\mathcal{F}_{\rm phot}$
 \item[] {(v)} Correct for colour-dependent basal flux $\mathcal{F}_{\rm min}$
 \item[] {(vi)} Calculate the excess flux $\Delta\mathcal{F}_{\rm Ca}$
 \end{enumerate}
 
The underlying idea here is that the flux in the cores of the Ca H and K lines (which we denote $\mathcal{F}_{\textup{1\angstrom}}$) contains the following components \citep[see][for discussion of this]{1989ApJ...341.1035S}:
 \begin{itemize}
 \item[]{--} Flux from photospheric line wings outside the H$_1$ and K$_1$ minima. In contrast to the chromosphere, the photosphere is assumed to be in radiative equilibrium. We denote this component $\mathcal{F}_{\rm phot}$.
  \item[]{--} Basal flux from a optically thick chromosphere that is unrelated to dynamo fields. We denote this component $\mathcal{F}_{\rm min}$.
  \item[]{--} Excess flux. This is the surface flux arising from magnetic sources -- i.e. a dynamo. This is the flux which we are interested in measuring. We denote this component $\Delta\mathcal{F}_{\rm Ca}$. 
 \end{itemize}
It follows that:
  \begin{equation}
 \Delta\mathcal{F}_{\rm Ca} = \mathcal{F}_{1\angstrom} - \mathcal{F}_{\rm phot} - \mathcal{F}_{\rm min}
 \end{equation}
 
\subsection{Correcting for blanketing}
We first correct the spectra
for line blanketing -- the decrease in intensity due to many closely 
spaced and thus unresolved lines. We use the blanketing coefficients 
from \citet{1995ApJ...438..404H}:
 \begin{equation}
 \epsilon'(3912)=1.032-0.296(B-V)
 \end{equation}
 \begin{equation}
 \epsilon'(4000)=1.060-0.167(B-V)
 \end{equation}
The correction is applied by making a linear fit through the two spectral points at $\lambda$3912 and $\lambda$4000 and then adjusting the 
values of this fit to $\epsilon'(3912)$ and $\epsilon'(4000)$.
 
\subsection{Normalising the spectra to an absolute flux scale}
In order to convert the spectra to absolute flux we use the absolute 
flux scale from \citet{2007AJ....133..862H}:
 \begin{equation}
 {\rm log} \mathcal{F}\left(\Delta\lambda\right)=8.179-2.887(b-y),-0.10 \leq b-y\leq 0.41
 \end{equation}
 \begin{equation}
 {\rm log} \mathcal{F}\left(\Delta\lambda\right)=8.906-4.659(b-y), 0.41 \leq b-y\leq 0.80
 \end{equation}
$\mathcal{F}\left(\Delta\lambda\right)$ is calculated as the 
flux density between $\lambda$3925 and $\lambda$3975 in units of ergs cm$^{-2}$ s$^{-1}$ {\AA}$^{-1}$.

The Str{\" o}mgren $b-y$ colour indices were calculated using the transformations from \citet{1996A&A...313..873A}:
 \begin{eqnarray}
 \theta_{\rm eff}=&&0.537+0.854(b-y)+0.196(b-y)^2\\
                   &-&0.198(b-y)c_1-0.026(b-y)\left[{\rm Fe/H}\right]\nonumber\\
                   &-&0.014\left[{\rm Fe/H}\right]-0.009\left[{\rm Fe/H}\right]^2\nonumber
 \end{eqnarray}
where $\theta_{\rm eff}=(5040 \, {\rm K})/T_{\rm eff}$ and assuming $c_1$=0.35. 
  
\subsection{Measuring $\mathcal{F}_{\textup{1\angstrom}}$}
The next step is to measure the integrated flux in a 1{\AA} bandpass 
$\mathcal{F}_{\textup{1\angstrom}}$ centred on the cores of the K and H 
lines. This is easily done in the wavelength corrected and velocity-zeroed 
spectra simply by summing the flux in the 1{\AA} bandpass.
  
\subsection{Correcting for photospheric flux $\mathcal{F}_{\rm phot}$}
The 1{\AA} bandpass flux $\mathcal{F}_{\textup{1\angstrom}}$ centred on 
the cores of the K and H lines will contain flux from the photosphere 
and from a colour-dependent basal flux, which could have an acoustic 
origin \citep[see, e.g., the review by][]{1995A&ARv...6..181S}.

In order to correct for the photospheric flux we need to calculate the 
separation $W_0$ between the two emission lines in the Ca cores 
\citep[see][]{2007AJ....133..862H}. This value can be calculated from 
\citet{1982MNRAS.199.1101L}:
 \begin{equation}
 {\rm log} W_0=-0.22 {\rm log} g +1.65 {\rm log} T_{\rm eff}+0.10[{\rm Fe/H}]-3.39
 \end{equation}
In Fig.~3 we have plotted $W_0$ as a function of effective temperature 
for the isochrones also used in Fig.~1.

We then calculate the flux from the photosphere by adjusting the 
integrated flux in the 1{\AA} bandpass 
($\mathcal{F}_{\textup{1\angstrom}}$) for the ratio between the fluxes in the $W_0$ 
and the 1{\AA} bandpass:
 \begin{equation}
 \frac{\mathcal{F}(W_0)}{\mathcal{F}(1\angstrom)}=W_0\mathcal{F}_{1\angstrom}^{1/4}
 \end{equation}
This ratio has also been corrected for the \F$^{1/4}$ activity 
scaling law from \citet{1979ApJ...228..509A}, and can be found in Fig.~4. Note that the flux in the $W_0$ bandpass is equivalent to the flux between the K$_1$ minima $\mathcal{F}({\rm K}_1)$ used by \citet{2007AJ....133..862H}.

\subsection{Correcting for colour-dependent basal flux $\mathcal{F}_{\rm min}$}
\citet{2007AJ....133..862H} obtained a relation between the Str{\" o}mgren $b-y$ colour indices and the photospheric contribution to the measured flux which can be used to correct these for the basal flux. We did try to use this relation along with $b-y$ colour indices calculated using the relations from \citet{1996A&A...313..873A}, but the resulting values for the basal flux were significantly overestimated resulting in negative values of \dF. We therefore adopted our own formulation of the basal flux -- estimated from the effective temperatures rather than the $b-y$ colour indices. This was done by making a linear fit to the mean fluxes in Fig.~5 and then lowering the fit values by  $3\sigma$ in order to get a representation of the lower level of the mean fluxes as function of effective temperature. The new formulation of the basal flux is:
 \begin{equation}
 {\rm log} \mathcal{F}_{\rm min}=3.07-5.11\cdot 10^{-4}T_{\rm eff}.
 \end{equation}
Note here that a number of different formulations exist in the literature for the basal flux \citep[see][for discussion of this]{1995A&ARv...6..181S}. The differences in these formulations most likely arise from slightly different instrument configurations and data reduction procedures, which also explains why we need to adopt our own formulation. We did test the effect on the final results of changing the formulation. This was done by varying the values of the lower level of the mean fluxes, which did not lead to any significant changes in the final results.

\subsection{Calculating the $S$ index}
The most commonly used expression for stellar chromospheric activity is the dimensionless $S$ index obtained from the spectrophotometers used in the Mount Wilson survey. Unfortunately, the $S$ index is sensitive to both the instrument configuration and the spectral resolution \citep{2007AJ....133..862H}. This is often accounted for by including a normalization constant $\alpha$. This constant is then calculated by observing a large number of stars from the Mount Wilson and Lowell surveys and using these stars as reference stars \citep[see e.g.][]{2004ApJS..152..261W}, but due to the intrinsic variability of the stars this number needs to be relatively large -- i.e. larger than our target list. 

We therefore adopted another approach and calculated the $S$ index 
as in \citet{2007AJ....133..862H} by measuring the flux in the H and K 
line cores using a 1.09 $\angstrom$ FWHM triangular filter as well as in 
two 20 $\angstrom$ reference bandpasses centred on $\lambda$3901.067 
($V$) and $\lambda$4001.067 ($R$).

The relationship between the $S$ index and 
$\mathcal{F}_{\textup{1\angstrom}}$ was then used to calibrate the $S$ index. This is 
done by calculating a pseudo $S$ index called $S'$ based on 
$\mathcal{F}_{\textup{1\angstrom}}$:
 \begin{equation}
 \mathcal{F}_{\textup{1\angstrom}}=10^{-14}S'C_{\rm cf}T_{\rm eff}^4,
 \end{equation}  
where $C_{\rm cf}$ is given as \citep{1984A&A...130..353R}:
 \begin{equation}
 {\rm log}\left(C_{\rm cf}\right)=0.45-0.066(B-V)^3-0.25(B-V)^2-0.49(B-V),
 \end{equation}
We then obtain a linear relation between the observed $S$ index called $S_{o}$ 
and the pseudo $S$ index ($S'$) which we can use to calibrate the $S$ index. In this way the observed 
$S$ index ($S_{o}$) is related to the $S$ index according to:
 \begin{equation}
 S=16.6S_{o}.
 \end{equation}
The constant of 16.6 is similar to the $\alpha$ calibration constant normally used to account for different spectral coverage and resolution when measuring the $S$ index with different instruments, and it normally lies between 1.8 and 5 \citep[see][]{2003AJ....126.2048G, 2004ApJS..152..261W, 2006AJ....132..161G, 2007AJ....133..862H}. The larger value found here is mainly due to the higher resolution of the NOT FIES spectra compared to other instruments that have been used to measure the $S$ index.

\subsection{Comparison to HD 157214}
The G0V star HD 157214 is at (17:20:39.30 +32:28:21.15) located close to 
the $Kepler$ field and this star is also part of the priority 1 list of 
stars being observed with the Solar-Stellar Spectrograph at Lowell 
Observatory. We have therefore observed this star together with the 
other stars on most observing nights. With the procedure described 
above we measure a mean excess flux of $4.8 \pm 0.2 \cdot 10^5 {\rm 
ergs}$ \rm ${\rm cm}^{-2} {\rm s}^{-1}$ and a mean $S$ index of $0.147 \pm 
0.004$ (where 0.004 is the uncertainty on the mean value). \citet{2007AJ....133..862H}
report a mean $S$ index of 0.162 and an excess flux that varies around 
$4.5 \cdot 10^5 {\rm ergs}$ ${\rm cm} ^{-2} {\rm s}^{-1}$ between 1995 
and 2007, but speculate that this star might be heading into an activity minimum. This speculation is supported by observations in 2008 and 2009, where the measured $S$ index of this star was 0.152 (Hall, private communication). Our mean $S$ index of $0.147 \pm 0.004$ is therefore in agreement with \citet{2007AJ....133..862H} -- given that the star has been observed at different activity phase and epoch.

The better agreement between the mean values of the excess flux, compared to the $S$ index, measured by \citet{2007AJ....133..862H} and us could reflect the fact that the external precision is much lower for the $S$ index than for the excess flux. In other words, the given instrument configuration is more important for the $S$ index than for the excess flux. This means that uncertainties in the $\alpha$ constant in eq. 12 lead to larger relative differences \citep[of the order of 10\%,][]{2007AJ....133..862H} in the $S$ index compared to the excess flux. Also, as the uncertainties we quote are calculated as the uncertainties of the mean values, they reflect only internal uncertainties and do not include offsets or biases between the observations presented here and the observations by e.g. \citet{2007AJ....133..862H}.

\subsection{Comparison to \citet{2010ApJ...725..875I}}
The $S$ index has also been measured for three stars in the sample by the California Planet Search program \citep{2010ApJ...725..875I}. They obtained one measurement of KIC~6116048 on 22 July 2010, 33 measurements of KIC~8006161 between 26 June 2005 and 4 June 2009 and 27 measurements of KIC~12258514 between 25 April 2010 and 15 September 2010. The mean values and the standard deviations of their and our measurements are provided in Table~3. As can been seen in the table the mean values agree within one standard deviation for KIC~6116048 and KIC~12258514. The $S$ index of KIC~8006161 has also been measured by \citet{1991ApJS...76..383D} who measured a mean $S$ index of 0.232 with a standard deviation of 0.004 between 2 May 1978 and 17 July 1978. Comparing the three mean values of the $S$ index of KIC~8006161 it appears that it has been declining between 1978 and now -- which could explain why we do not have agreement within one standard deviation between the mean value by \citet{2010ApJ...725..875I} and the mean value by us.

\subsection{Rotation Periods}
Stellar rotation periods can be measured in the white-light observations from {\it Kepler} simply by calculating a periodogram and identifying the highest peak in this periodogram \citep[see e.g.][]{2012A&A...539A.137M, 2012AN....333.1036N, 2013arXiv1305.5721N}. Though this method is simple, care is needed since the method can be susceptible to bias, e.g. the highest peak could be the second or third harmonic of the true rotation period. Part of this problem can be solved by comparing the estimated rotation periods to independent estimates from e.g. $v$sin$i$ measurements and asteroseismology. Results of such a analysis will be presented elsewhere (Garc{\'{\i}}a et al., in preparation.). Here we present rotation periods for only a limited number of stars for which rotation periods could be unambiguously identified; even these periods may change slightly when additional {\it Kepler} observations become available.

We searched for rotational modulations of the light curves by 
calculating a least-square periodogram (Lomb 1976, see also Karoff 2008). 
The periodograms were calculated from 14 quarters of $Kepler$ 
observations that were gathered between 13 May 2009 and 3 October 2012.
For this part of the analysis we only used long-cadence 
observations because not all short-cadence observations were available in a 
PDC (Pre-search Data Conditioning) processed format. The higher 
time-resolution of the short-cadence data is only needed for the
asteroseismic analysis and not for searching for rotational modulations 
in the light curves. The PDC processed observations 
have a significantly reduced number of artifacts that could mimic a spot on the 
surface of the stars compared to the Simple Aperture Photometry (SAP) generated 
by the PA (Photometric Analysis) pipeline module.

The periodograms were calculated for 1000 periods between 0 and 20 days 
in each of the 14 quarters of data for each star. We then identified any peaks in 
these periodograms with a S/N higher than 4 and crosschecked these peaks 
with the signal in the light curves. For a star to be assigned a rotation period, the same peak was required to be visible in the 
periodograms during all 14 quarters and in the light curves. 

To validate these rotation periods we compare them in Fig.~6
with the values one would obtain from the 
$v$sin$i$ measurements by Bruntt et al. (2012; see Table~2) and the 
asteroseismic radii (Table~2). Indication of a linear trend is 
seen in the figure, but most of the data points fall above 
the linear relation indicated by the solid line. This is expected since: 1) 
sin$i$ will take values between zero and one, 2) it seems natural to 
expect spots to form close to the fastest rotation latitudes on the 
stellar surfaces, and 3) there are inherent uncertainties in the $v$sin$i$ measurements -- such as choosing the macroturbulence parameter, etc.

\citet{2010ApJ...725..875I} detected a 19-day rotation period in KIC 12258514, which is consistent with the fact that we do not see any significant peaks in our periodograms calculated between 0 and 20 days. They also detected a 43-day period in KIC 8006161. Originally we did detect a 10-day period in the periodograms of KIC 8006161 calculated as described above. In order to solve this incongruence we calculated new periodograms for 3000 periods between 0 and 60 days. The periodograms revealed that KIC 8006161 shows longer more prominent periods than the 10-day period. None of these periods could meet the criteria described above and no rotation period was thus assigned to KIC 8006161. The same phenomenon was seen in KIC 6116048.

In total we were able to assign a rotation period to 10 out of the 19 stars. Periodograms for these 10 stars are shown in Appendix 1. We note that some of the remaining stars are likely to have rotation periods longer than 20 days.

\section{Results}
We have measured both the $S$ index and the excess flux 
$\Delta\mathcal{F}_{\rm Ca}$ for each of our stars (see Table~2). 
The main difference between 
these two quantities is that different colour-dependent terms have been 
removed from the excess flux $\Delta\mathcal{F}_{\rm Ca}$ 
\citep{2007AJ....133..862H}. These terms become important when comparing 
the measured activity level in the stars to stellar properties. Here the 
excess flux is the best quantity to use because it is corrected for the 
different colour-dependent terms. Another way to explain the difference 
between the $S$ index and the excess flux is that the $S$ index is a 
relative measurement of the activity in the stars -- relative in the 
sense that it measures the intensity in the $H$ and $K$ bandpasses relative 
to the $R$ and $V$ bandpasses. The excess flux, on the other hand, is an 
absolute measurement of the activity of the stars -- corrected for 
terms that are not related to (magnetic) activity.

All of the results we present here are mean values of the measured 
quantities in the 8 different epochs that have been observed so far, and 
the error bars represent the uncertainties on the mean values.
Fig.~7 compares the measured mean $S$ index to the mean excess flux. A 
clear log-linear relation is seen for all 19 stars except one, which is KIC 
8006161. KIC 8006161 has a mean $S$ index of $0.172 \pm 0.002$, 
which is comparable to the mean value seen in the Sun, but the excess flux is only 
$9.8 \pm 1.4 \cdot 10^4$ ergs cm$^{-2}$ s$^{-1}$, which is close to the quiet Sun. 
A handful of stars showing a similar behaviour (not obeying the linear 
relation between the $S$ index and the excess flux and showing lower 
than expected excess flux) were identified by 
\citet{2007AJ....133..862H} -- the most prominent being $\tau$~Cet.  
$\tau$~Cet has long been suspected of being a Maunder minimum star 
\citep{2004ApJ...609..392J} and it will therefore be interesting to see 
whether KIC 8006161 also shows low variability in the excess flux.

We have also analysed the relationship between the relative variability 
of the $S$ index and the excess flux and the mean value of these 
parameters (Figs.~8 \& 9). The relation for the 
excess flux generally follows the same trend that was seen by 
\citet{2007AJ....133..862H}.

\subsection{Activity distributions}
\citet{1980PASP...92..385V} were the first to note an apparent 
deficit in the number of F-G stars exhibiting intermediate activity. 
This gap, which is now known as the Vaughan-Preston gap, has been studied 
extensively since then. One of the largest studies of the gap was performed by 
\citet{1996AJ....111..439H}, who showed that two Gaussian functions were 
needed to satisfactorily model the activity distribution of more 
than 800 southern stars within 50 pc.

To evaluate how the stars in this study are distributed around the 
Vaughan-Preston gap, we have calculated the distributions for both the $S$ 
index and the excess flux in Figs.~10 \& 11. The bimodal distribution in 
stellar activity cannot be clearly identified in either of the two histograms.
This is in agreement with the results from \citet{2007AJ....133..862H} 
who were also unable to find clear indications of a bimodal distribution (Fig.~12 in their paper). In 
fact our distribution looks almost identical to the distribution of 
\citet{2007AJ....133..862H}, although it should be borne in mind that they have 143 stars and we have only 19. A Kolmogorov-Smirnov test comparing the two distributions yields a p-value of 0.78, supporting the contention that the measurements arise from the same underlying distributions.

If simple counting statistics were adopted for the uncertainties of each bin, it would, on the other hand, become clear that we would not be able to see a bimodal distribution, even if it were intrinsic here as we only have 19 stars in our sample. 

\subsection{Age-Rotation-Activity Relations}
It was suggested by \citet{1972ApJ...171..565S} that there exists a power-law relation between rotation and activity on the one side and age on the other. This study forms the basis for an assumption of a causal relationship between age, rotation and activity -- the so-called age-rotation-activity relations \citep[see e.g.][or the discussion in the introduction]{2001ApJ...563..334S, 2007ApJ...669.1167B}.
However, as discussed in the introduction, 
the number of data points used for calculating these relations leaves 
room for improvement -- especially for stars of solar age or older. The 
10 stars for which we have measured rotation periods provide such an 
improvement.

Despite the lack of Sun-like stars with independently measured ages that can be used 
to improve the relations by \citet{1972ApJ...171..565S}, much work has 
gone into improving the theoretical understanding of these relations 
\cite[see e.g.][]{1988ApJ...333..236K, 1990ApJS...74..501P} and it is clear that no simple log-linear relation between age, 
rotation and activity exist in general for F-G main sequence stars \citep[see e.g.][]{1984ApJ...279..763N, 1991ApJ...375..722S, 1998ASPC..154.1235D, 2008ApJ...687.1264M}. In 
other words, when more data than the four points used by
\citet{1972ApJ...171..565S} are available, it is not obvious that all 
the new data points should follow the simple log-linear relations. 
Nevertheless, we have fitted all of our measured age-rotation-activity 
relations with log-linear fits for illustrative purposes.

The first relation we have looked at is the relation between the 
rotation period and the excess flux in Fig.~12. A log-linear relation is 
clearly seen in the figure, and is represented by the solid line given by:
 \begin{equation}
 \log \Delta\mathcal{F}_{\rm Ca}=(-0.74\pm0.03)\log P_{\rm rot}+6.55\pm0.02
 \end{equation}
The exponent of $-0.74\pm0.03$ is in agreement with the 
original result by \citet{1972ApJ...171..565S}.

The second relation is between the age and the excess flux in Fig.~13. 
The log-linear relation found here is given by:
 \begin{equation}
\log \Delta\mathcal{F}_{\rm Ca}=(-0.61\pm0.17)\log {\rm Age}+6.08\pm0.13
 \end{equation}
The important parameter to compare here is the exponent $-0.61\pm0.17$. 
\citet{1972ApJ...171..565S} found this value to be $-0.54$ and 
\citet{1991ApJ...375..722S} found it to be $-0.66$ (though for $R'_{\rm 
HK}$ instead of $\Delta\mathcal{F}_{\rm Ca}$). Our result does in other words agree with both results within $1\sigma$.

The last relation to analyse is between the rotation period and the age 
in Fig.~14. Here the log-linear relation is given by: 
 \begin{equation} 
 \log P_{\rm rot}=(0.45\pm0.19) \log {\rm Age}+0.59\pm0.29
 \end{equation} 
The exponent of $0.45\pm0.19$ compares nicely to the value of $0.51$ 
found by \citet{1972ApJ...171..565S}.

It has been shown by e.g. \citet{2003ApJ...586..464B} that the rotation period is not only a function of age, but also of colour (or, equivalent, mass). The reason for this is likely that different spin-down time scales exist for stars of different masses. We have therefore included a $B-V$ colour term in the model. This was done in a way similar to that demonstrated by \citet{2003ApJ...586..464B}, although in the function $f$ we have replaced the value 0.50 used by Barnes to 0.38, to account for the range in $B-V$ of the 10 stars in this study (the different offsets does not have any other implications). This provided us with the following relation between rotation, age and $B-V$ colour (see Fig.~15):

\begin{equation}
\log P_{\rm rot}=(0.81\pm0.10) \log {\rm Age}+ \log f(B-V)+0.47\pm0.22,
\end{equation}
 where
 \begin{equation}
f(B-V)=\sqrt{B-V-0.38}-0.15(B-V-0.38).
\end{equation}

The reduced $\chi^2$ value of this fit was 1.46, which is lower than the value of 2.57, which we obtain if we use the Skumanich model (eq. 15).

\section{Discussion}
We have chosen to measure both the excess flux and the $S$ index. This was done as it was the only
way to calibrate our measured $S$ indices. Validations of the results presented here have, on the other hand, also shown us that we can obtain a much stronger 
relation between stellar properties measured with asteroseismology 
and measured values of the excess flux than with measured values of the $S$ index. This 
strengthens the general proposition that the $S$ index includes a number of 
(colour dependent) terms that do not relate to the evolution of stellar 
angular momentum and activity \citep{1982A&A...107...31M, 1984A&A...130..353R, 1987A&A...177..155R, 2007AJ....133..862H}.

\subsection{Do we cover both sides of the Vaughan-Preston gap?}
One of the main goals of the target selection was to ensure that the 
selected targets would cover both sides of the Vaughan-Preston gap. From 
Fig.~11 it seems that we have only partly succeeded in this -- all the 
stars appear to fall on the inactive branch with $S$ indices less than 0.2. 
Of course a histogram made from 19 data points must be taken with 
caution, and some of the 19 stars apparently on the inactive branch might 
turn out to be active stars. This is also reflected in Fig.~10 where we 
see that our distribution of 19 stars is in agreement with the 
distribution of 143 obtained by \citet{2007AJ....133..862H}. This 
suggests that we are sampling typical excess fluxes for Sun-like stars 
and that our sample also includes six stars on the active sequence with 
log$\Delta\mathcal{F}_{\rm Ca}$ above 6 (contradicting what is seen in the results based on the $S$ index). The asteroseismic ages of the stars also suggest that we do cover stars on both sides of the gap.

\subsection{What can we learn from the Age-Rotation-Activity Relations?}
The 10 stars for which we have independent measurements of asteroseismic ages, rotation periods and excess flux generally all fulfill the Skumanich relations. The possible exception here is the Sun, whose rotation rate and excess flux seem to be significantly lower than predicted by the Skumanich relation, KIC 8006161 (star i in Table~2), whose excess flux also seems to be lower than predicted by the Skumanich relation, and KIC 11244118 (star p in Table~2), whose rotation period seems longer than predicted by the Skumanich relation. A possible explanation for the low excess flux of KIC 8006161 could be that it is in a Maunder minimum state. The long rotation period of KIC 11244118 could be related to the fact that, with an age of 6.3$_{-4.3}^{+1.2}$ Gyr, it is a relatively old star considering its mass of 1.23$^{+0.10}_{-0.08}$ M$_{\sun}$; in other words this star is likely the most evolved star in the sample and might not be a main-sequence star, but a sub-giant (which is also in agreement with its location in Fig.~1).

\citet{2003ApJ...586..464B} suggested that the relation between stellar rotation period and age separated into two sequences -- one for Sun-like stars called the interface sequence and one for the younger G, K and M dwarfs called the convective sequence. As expected we are not able to identify this bifunctionality in the 10 stars analysed here. The reason for this is likely that none of the 10 stars are so young that they fall on the convective sequence.

\section*{Acknowledgments}
This work was partially supported by NASA grant NNX13AC44G. Funding for this Discovery mission is provided by NASA's Science Mission 
Directorate. The authors wish to thank the entire {\it Kepler} team, without 
whom these results would not be possible. We also thank all funding 
councils and agencies that have supported the activities of KASC Working 
Group 1, and the International Space Science Institute (ISSI). CK 
acknowledges support from the Carlsberg foundation. WJC acknowledges the support of the UK Science and Technology Facilities Council (STFC). Funding for the Stellar Astrophysics Centre is provided by The Danish National Research Foundation (Grant DNRF106). The research is supported by the ASTERISK project (ASTERoseismic Investigations with SONG and {\it Kepler}) funded by the European Research Council (Grant agreement no.: 267864).


\clearpage

\begin{table}
\caption{Target list for the {\it sounding stellar cycles with Kepler} programme. We also list  the Kepler magnitude, $B-V$ values from \citet{2000A&A...355L..27H} and Hipparcos luminosities in units of solar luminosities from The Hipparcos and Tycho Catalogues (ESA 1997) for the the Hipparcos stars in the sample.}
\centering
\begin{tabular}{lccccccc}
\hline \hline
KIC ID & $\alpha$ (2000) & $\delta$ (2000) & $k_p$ & $B-V$ & $L$\\
\hline
01435467 & 19:28:19.84 & 37:03:35.3 & 8.9 & 0.47$\pm$0.02 &\\ 
02837475 & 19:10:11.62 & 38:04:55.9 & 8.4 & 0.43$\pm$0.02 &\\
03733735 & 19:09:01.92 & 38:53:59.6 & 8.4 & 0.41$\pm$0.02 & 3.79$\pm$0.49 \\
04914923 & 19:16:34.88 & 40:02:50.1 & 9.4 & 0.62$\pm$0.03 & 2.32$\pm$0.58\\
06116048 & 19:17:46.34 & 41:24:36.6 & 8.4 & 0.57$\pm$0.01 &\\ 
06603624 & 19:24:11.18 & 42:03:09.7 & 9.0 & 0.76$\pm$0.03 &\\
06933899 & 19:06:58.34 & 42:26:08.2 & 9.6 & 0.59$\pm$0.04 &\\
07206837 & 19:35:03.72 & 42:44:16.5 & 9.7 & 0.46$\pm$0.06 &\\
08006161 & 18:44:35.14 & 43:49:59.9 & 7.3 & 0.87$\pm$0.01 & 0.61$\pm$0.02 \\
08379927 & 19:46:41.28 & 44:20:54.7 & 6.9 & 0.58$\pm$0.01 & 1.05$\pm$0.08\\
08694723 & 19:35:50.58 & 44:52:49.8 & 8.8 & 0.48$\pm$0.02 &\\
09098294 & 19:40:21.20 & 45:29:20.9 & 9.7 & 0.68$\pm$0.08 &\\
09139151 & 18:56:21.26 & 45:30:53.1 & 9.1 & 0.52$\pm$0.03 & 1.63$\pm$0.40\\
09139163 & 18:56:22.12 & 45:30:25.2 & 8.3 & 0.49$\pm$0.01 & 3.88$\pm$0.69\\
10124866 & 18:58:03.46 & 47:11:29.9 & 7.9 & 0.57$\pm$0.02 &\\  
10454113 & 18:56:36.62 & 47:39:23.0 & 8.6 & 0.52$\pm$0.02 & 2.60$\pm$0.36\\
11244118 & 19:27:20.48 & 48:57:12.1 & 9.7 & 0.78$\pm$0.05 &\\
11253226 & 19:43:39.62 & 48:55:44.2 & 8.4 & 0.39$\pm$0.02 & 4.22$\pm$0.61\\
12009504 & 19:17:45.80 & 50:28:48.2 & 9.3 & 0.55$\pm$0.03 &\\
12258514 & 19:26:22.06 & 50:59:14.0 & 8.0 & 0.59$\pm$0.01 & 2.84$\pm$0.25\\
\hline
\end{tabular}
\label{tab1}
\end{table}

\begin{table*}
\caption{Measured stellar properties.  The short names refer to the symbols used in Figs.~2, 8 \& 9. The stellar parameters are measured with asteroseismology. $T_{\rm eff}$ and $v$sin$i$ values are from \citet{2012MNRAS.423..122B}. $P_{\rm rot}$, log $<\Delta \mathcal{F}_{Ca}>$ and $<S>$ values are from this study.}
\centering
\rotatebox{90}{
\begin{tabular}{lccccccccccc}
\hline \hline
KIC 		& 	Short& 	$T_{\rm eff}$ [K]		& 	[Fe/H] 			& log $g$						& 	$M$[M$_{\odot}$] 		&	$R$ [R$_{\odot}$] 		& age [Gyr] 			& $v$sin$i$ [km/sec] & $P_{\rm rot}$ [days] 		& log $<\Delta \mathcal{F}_{\rm Ca}>$ 	& $<S>$ \\
\hline
01435467 &	a	&       6222$\pm$60		&	-0.01$\pm$0.06	&	4.077$^{+0.020}_{-0.016}$ 	&	1.22$^{+0.10}_{-0.09}$ 	&	1.66$^{+0.04}_{-0.04}$ 	&	4.2$^{+1.2}_{-1.5}$ 	&	10.0			&	7.2 $\pm$     0.3	&	5.97 $\pm$    0.01 			&      0.157 $\pm$   0.001\\
02837475 &	b	&	6741$\pm$60		&	-0.02$\pm$0.06	&	4.155$^{+0.026}_{-0.026}$ 	&	1.36$^{+0.06}_{-0.10}$ 	&	1.60$^{+0.04}_{-0.04}$ 	&	2.0$^{+0.9}_{-0.8}$ 	&	23.5 			&       3.7 $\pm$     0.1	&	6.13 $\pm$     0.01 			&      0.166 $\pm$   0.001\\
03733735 &	c	&	6687$\pm$60		&	-0.04$\pm$0.06	&	4.268$^{+0.015}_{-0.016}$ 	&	1.30$^{+0.07}_{-0.05}$ 	&	1.39$^{+0.03}_{-0.03}$ 	&	1.0$^{+1.0}_{-0.7}$ 	&	16.8 			&       2.6 $\pm$     0.1	&	6.22 $\pm$    0.01 			&      0.182 $\pm$   0.001\\
04914923 &	d	&       5798$\pm$60		&	0.17$\pm$0.06		&	4.198$^{+0.015}_{-0.016}$ 	&	1.11$^{+0.10}_{-0.08}$ 	&	1.39$^{+0.04}_{-0.04}$ 	&	7.6$^{+2.6}_{-3.1}$ 	&	3.6 			&       8.1 $\pm$      0.4	&	5.46 $\pm$     0.12 			&      0.137 $\pm$   0.005\\
06116048 &	e	&       6022$\pm$60		&	-0.24$\pm$0.06	&	4.250$^{+0.009}_{-0.011}$ 	&	0.92$^{+0.04}_{-0.06}$ 	&	1.19$^{+0.02}_{-0.02}$ 	&	8.9$^{+2.1}_{-1.9}$ 	&	4.0 			&       --				&	5.70 $\pm$     0.02 			&      0.152 $\pm$   0.001\\
06603624 &	f	&       5673$\pm$60		&	0.28$\pm$0.06		&	4.316$^{+0.008}_{-0.007}$ 	&	1.00$^{+0.03}_{-0.04}$ 	&	1.15$^{+0.01}_{-0.02}$ 	&	11.1$^{+1.4}_{-1.5}$ &	3.0 			&       --				&	5.47 $\pm$     0.07 			&      0.155 $\pm$   0.004\\
06933899 &	g	&       5907$\pm$60		&	0.02$\pm$0.06		&	4.091$^{+0.014}_{-0.014}$ 	&	1.15$^{+0.07}_{-0.07}$ 	&	1.59$^{+0.03}_{-0.03}$ 	&	4.9$^{+2.3}_{-1.6}$ 	&	3.5 			&       --				&	5.70 $\pm$     0.05 			&      0.149 $\pm$   0.003\\
07206837 &	h	&       6343$\pm$60		&	0.14$\pm$0.06		&	4.169$^{+0.026}_{-0.026}$ 	&	1.34$^{+0.08}_{-0.13}$ 	&	1.57$^{+0.05}_{-0.05}$ 	&	2.9$^{+2.1}_{-1.8}$ 	&	10.1 			&       --				&	5.67 $\pm$     0.05 			&      0.138 $\pm$   0.002\\
08006161 &	i	&       5291$\pm$60		&	0.34$\pm$0.06		&	4.490$^{+0.002}_{-0.001}$ 	&	0.97$^{+0.02}_{-0.01}$ 	&	0.92$^{+0.01}_{-0.01}$ 	&	5.2$^{+1.4}_{-0.1}$ &	2.5 			&       --				&	4.99 $\pm$     0.06 			&      0.172 $\pm$   0.002\\
08379927 &	j	&       6241$\pm$150	&	-0.10$\pm$0.10	&	4.373$^{+0.011}_{-0.010}$ 	&	1.03$^{+0.08}_{-0.05}$ 	&	1.09$^{+0.02}_{-0.02}$ 	&	2.5$^{+1.4}_{-1.1}$ &	--			&         -- 					&	5.99 $\pm$    0.01 			&      0.181 $\pm$   0.001\\
08694723 &	k	&       6287$\pm$60		&	-0.59$\pm$0.06	&	4.079$^{+0.035}_{-0.018}$ 	&	0.94$^{+0.08}_{-0.08}$ 	&	1.45$^{+0.02}_{-0.03}$ 	&	7.7$^{+2.4}_{-3.0}$ &	6.6 			&       7.5 $\pm$      0.2	&	5.78 $\pm$     0.03 			&      0.159 $\pm$   0.002\\
09098294 &	l	&       5830$\pm$60		&	-0.13$\pm$0.06	&	4.301$^{+0.012}_{-0.015}$ 	&	0.98$^{+0.05}_{-0.09}$ 	&	1.15$^{+0.02}_{-0.04}$ 	&	6.3$^{+4.8}_{-2.4}$ &	4.0 			&        --				&	5.64 $\pm$     0.04 			&      0.150 $\pm$   0.003 \\
09139151 &	m	&       6127$\pm$60		&	0.11$\pm$0.06		&	4.374$^{+0.012}_{-0.013}$ 	&	1.15$^{+0.06}_{-0.06}$ 	&	1.15$^{+0.03}_{-0.03}$ 	&	2.9$^{+2.1}_{-1.9}$ &	6.0 			&       10.4 $\pm$      0.4	&	5.85 $\pm$     0.02 			&      0.155 $\pm$   0.002\\
09139163 &	n	&       6341$\pm$60		&	0.15$\pm$0.06		&	4.193$^{+0.020}_{-0.022}$ 	&	1.35$^{+0.10}_{-0.11}$ 	&	1.54$^{+0.03}_{-0.03}$ 	&	2.5$^{+2.0}_{-1.7}$ &	4.0			&       6.5 $\pm$ 0.2		&	5.70 $\pm$     0.02 			&      0.143 $\pm$   0.001\\
10454113 &	o	&       6295$\pm$60		&	-0.06$\pm$0.06	&	4.304$^{+0.010}_{-0.010}$ 	&	1.11$^{+0.05}_{-0.05}$ 	&	1.23$^{+0.02}_{-0.02}$ 	&	2.0$^{+1.5}_{-0.9}$ &	5.5			&       --				&	5.94 $\pm$     0.01 			&      0.169 $\pm$   0.001\\
11244118 &	p	&       5590$\pm$60		&	0.35$\pm$0.06		&	4.092$^{+0.027}_{-0.011}$ 	&	1.23$^{+0.10}_{-0.08}$ 	&	1.64$^{+0.03}_{-0.03}$ 	&	6.3$^{+1.2}_{-4.3}$ &	3.0			&       18.7 $\pm$       2.4	&	5.66 $\pm$     0.04 			&      0.140 $\pm$   0.002\\
11253226 &	q	&       6520$\pm$60		&	-0.08$\pm$0.06	&	4.153$^{+0.030}_{-0.029}$ 	&	1.28$^{+0.08}_{-0.11}$ 	&	1.56$^{+0.05}_{-0.05}$ 	&	2.4$^{+1.3}_{-0.9}$ &	15.1 			&       3.8 $\pm$     0.1	&	6.23 $\pm$    0.01 			&      0.184 $\pm$   0.001 \\
12009504 &	r	&       6082$\pm$60		&	-0.09$\pm$0.06	&	4.194$^{+0.016}_{-0.015}$ 	&	1.03$^{+0.09}_{-0.07}$ 	&	1.35$^{+0.03}_{-0.03}$ 	&	6.5$^{+3.0}_{-2.9}$ &	8.4			&       9.6 $\pm$       1.1	&	5.84 $\pm$    0.01 			&      0.155 $\pm$   0.001\\
12258514 &	s	&       5935$\pm$60		&	0.04$\pm$0.06		&	4.102$^{+0.023}_{-0.013}$ 	&	1.15$^{+0.11}_{-0.09}$ 	&	1.57$^{+0.04}_{-0.03}$ 	&	5.9$^{+1.6}_{-2.6}$ &	3.5	 		&       --				&	5.83 $\pm$     0.02 			&      0.152 $\pm$   0.001\\
\hline
\end{tabular}}
\label{tab2}
\end{table*}

\clearpage

\begin{table}
\caption{Comparison between the $S$ indexes measured by \citet{2010ApJ...725..875I} and this study. Numbers in brackets are the standard deviation.}
\centering
\begin{tabular}{lcc}
\hline \hline
KIC ID & \citet{2010ApJ...725..875I} & This study\\
\hline
06116048 & 0.157 & 0.152 (0.007)\\
08006161 & 0.190 (0.006) & 0.172 (0.011)\\
12258514 & 0.158 (0.002) & 0.152 (0.010) \\	
\hline
\end{tabular}
\label{tab3}
\end{table}

\begin{figure}
\includegraphics[width=\columnwidth]{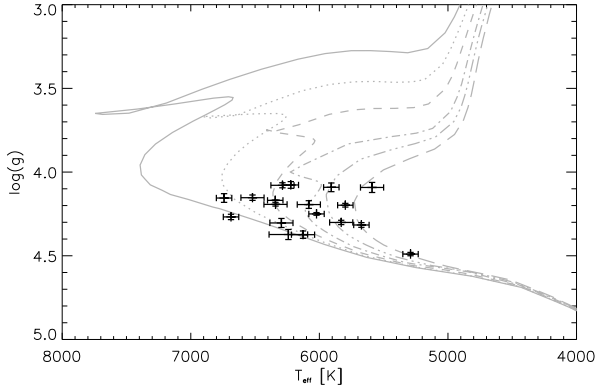}
\caption{The 20 stars in the programme along with Padova isochrones 
\citep{2004A&A...415..571B, 2002A&A...391..195G, 2004A&A...422..205G}
calculated for 6 different ages between 1 and 
10 Gyr in steps of 0.2 dex, using a metallicity of $Z$ = 0.02.}
\end{figure}

\begin{figure}
\includegraphics[width=\columnwidth]{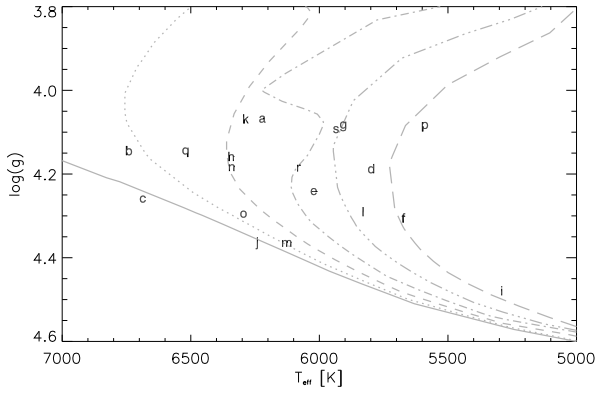}
\caption{A zoom in on figure 1 with the indvidual stars marked with unique letters.}
\end{figure}

\begin{figure}
\includegraphics[width=\columnwidth]{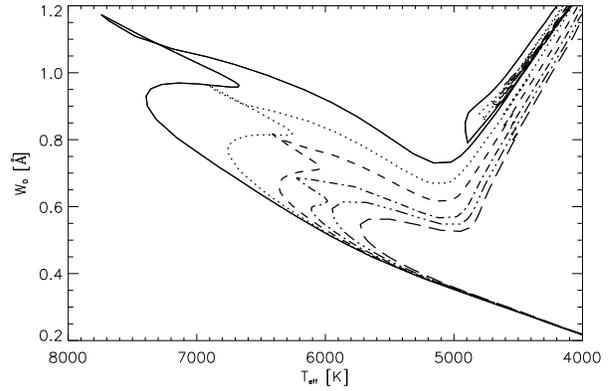}
\caption{Separation between the two emission lines in the Ca K$_2$ core 
of the K and H lines ($W_0$) as a function of effective temperature \citep[see e.g.][for definition of this separation]{1979ApJ...228..509A}. The 
isochrone tracks are the same as in Fig.~1.}
\end{figure}

\begin{figure}
\includegraphics[width=\columnwidth]{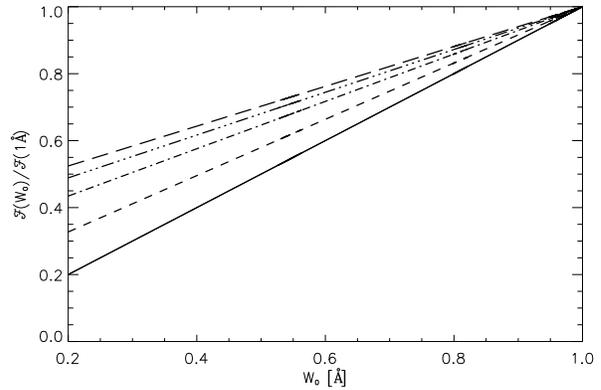}
\caption{Fraction of the flux in the 1$\angstrom$ bandpass lying within  the core of the K and H lines for stars with solar activity levels 
(bottom line) and up to 8 times solar level.}
\end{figure}

\begin{figure}
\includegraphics[width=\columnwidth]{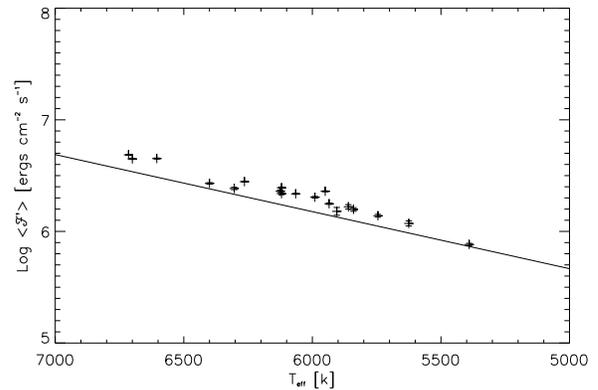}
\caption{Measured mean excess fluxes corrected for contribution from the 
photosphere plotted as function of effective temperature. The solid line shows 
the relation for the basal flux used in this study.} 
\end{figure}

\begin{figure}
\includegraphics[width=\columnwidth]{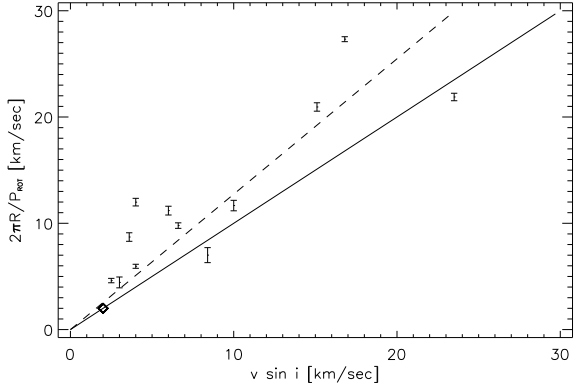}
\caption{Relation between the estimated equatorial velocities measured in the $Kepler$ 
light curves and the $v\sin i$ values from \citet{2012MNRAS.423..122B}. 
The rotation periods have been scaled with the asteroseismic radii in 
order to put the two quantities on the same scale. The solid line shows 
a 1:1 relation. As expected some points are seen above the line. 
This is partly due to the fact that stars come with random inclinations 
and partly due to the fact that stars have differential rotation (see 
text). The dashed line illustrates this by showing the relation for a most probable $v\sin i$ value of $\pi/4$ \citep{1992oasp.book.....G}. The diamond represents the Sun.} 
\end{figure}

\begin{figure}
\includegraphics[width=\columnwidth]{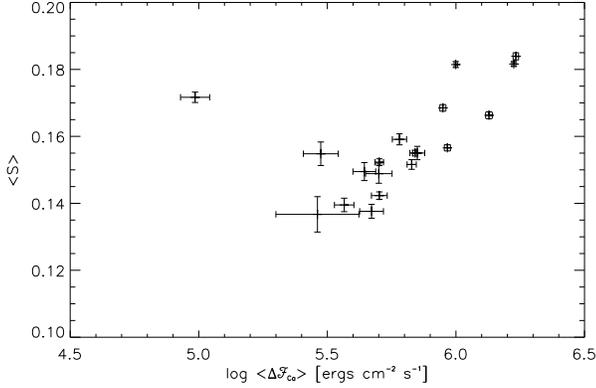}
\caption{The relation between the $S$ index and the excess flux 
$\Delta\mathcal{F}_{\rm Ca}$. A nice log-linear relation is seen between 
the two quantities except for KIC 8006161.} 
\end{figure}

\begin{figure}
\includegraphics[width=\columnwidth]{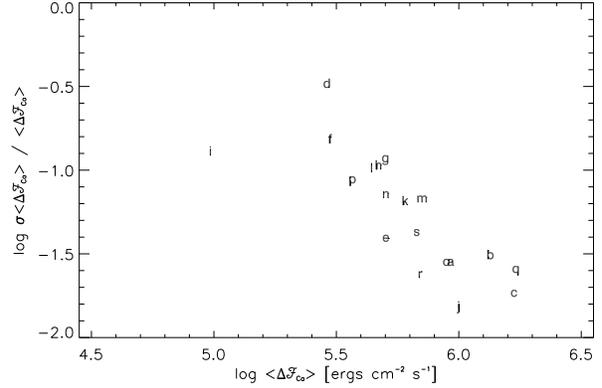}
\caption{The relative variability of the excess flux as a function of 
the mean value of the magnetic emission. } 
\end{figure}

\clearpage

\begin{figure}
\includegraphics[width=\columnwidth]{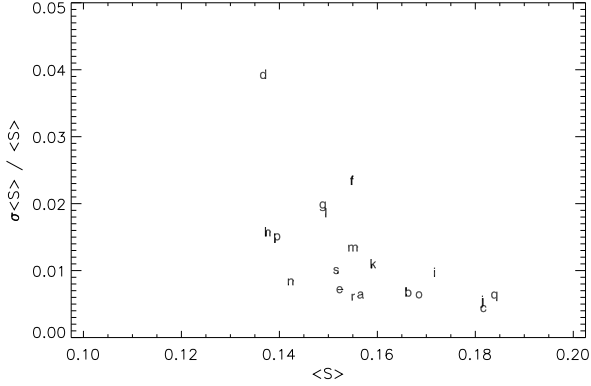}
\caption{The relative variability of the excess flux as a function of 
the mean value of the $S$ index. } 
\end{figure}

\begin{figure}
\includegraphics[width=\columnwidth]{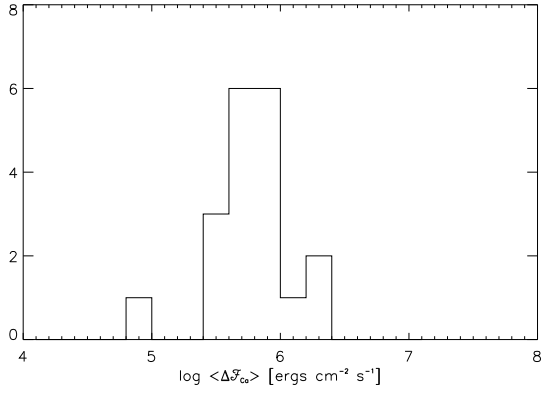}
\caption{Histogram of the measured mean magnetic emissions.}
\end{figure}

\begin{figure}
\includegraphics[width=\columnwidth]{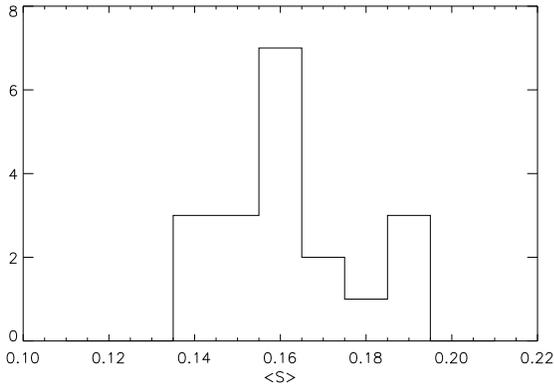}
\caption{Histogram of the measured mean $S$ indicies.}
\end{figure}

\begin{figure}
\includegraphics[width=\columnwidth]{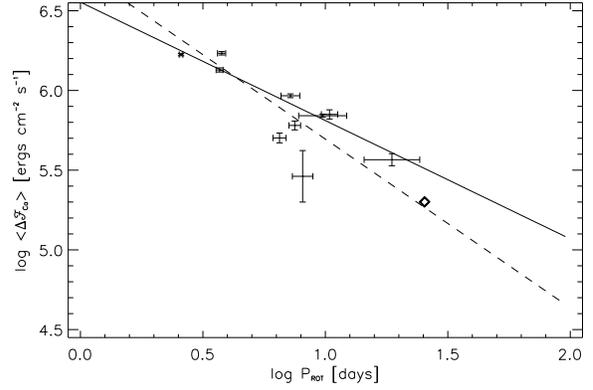}
\caption{Measured mean excess flux as a function of rotation period. The 
solid line gives the following log-linear relation: $\log \Delta\mathcal{F}_{\rm Ca}=(-0.74\pm0.03) \log P_{\rm rot}+6.55\pm0.02$. The diamond represents the Sun and the dashed line shows the log-linear relation from \citet{1972ApJ...171..565S}.}
\end{figure}

\begin{figure}
\includegraphics[width=\columnwidth]{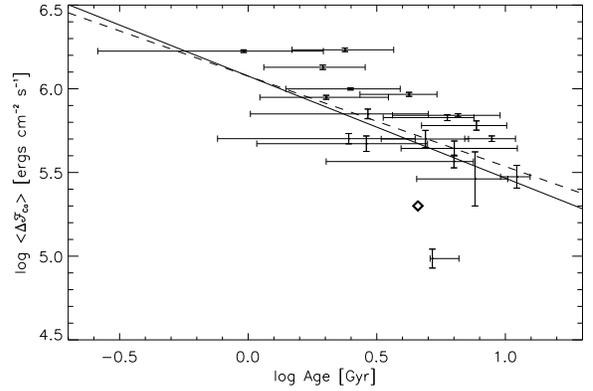}
\caption{Measured mean excess flux as a function of stellar age. The 
solid line gives the following log-linear relation: $\log \Delta\mathcal{F}_{\rm Ca}=(-0.61\pm0.17) \log {\rm Age}+6.08\pm0.13$. 
The diamond represents the Sun and the dashed line shows the log-linear relation from \citet{1972ApJ...171..565S}.} 
\end{figure}

\clearpage

\begin{figure}
\includegraphics[width=\columnwidth]{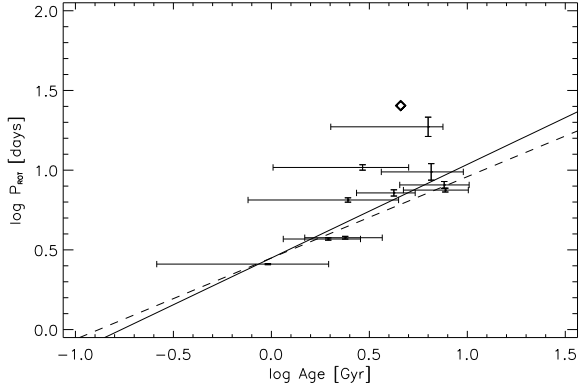}
\caption{Rotation period as a function of age. The solid line gives the 
following log-linear relation: $\log P_{\rm rot}=(0.45\pm0.19) \log {\rm Age}+0.59\pm0.29$. The diamond represents the Sun and the dashed line shows the log-linear relation from \citet{1972ApJ...171..565S}. The reduced $\chi^2$ value of the fit is 2.57.} 
\end{figure}

\begin{figure}
\includegraphics[width=\columnwidth]{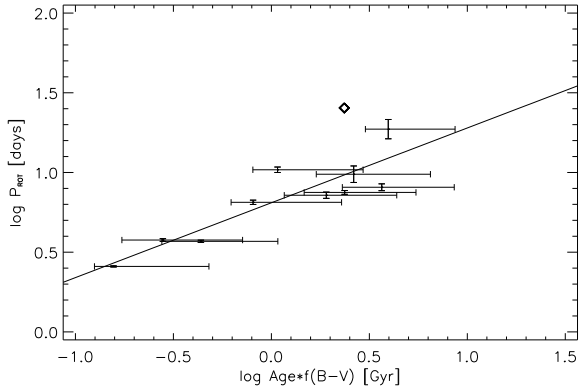}
\caption{Rotation period as a function of age and $B-V$ colour. The solid line gives the 
following log-linear relation: $\log P_{\rm rot}=(0.81\pm0.10) \log {\rm Age}+ \log f(B-V)+0.47\pm0.22$. The diamond represents the Sun. The reduced $\chi^2$ value of the fit is 1.46 -- indicating that the colour term in the model is real.} 
\end{figure}

\appendix

\begin{center}
\begin{figure*}
\centerline{\hbox{\includegraphics[width=6.0cm]{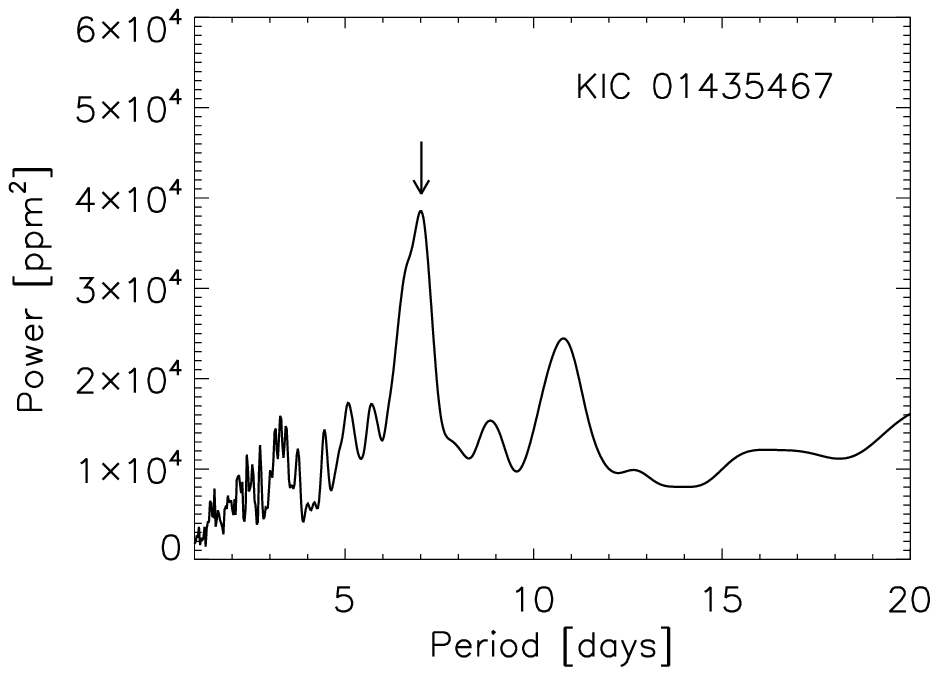}  
\includegraphics[width=6.0cm]{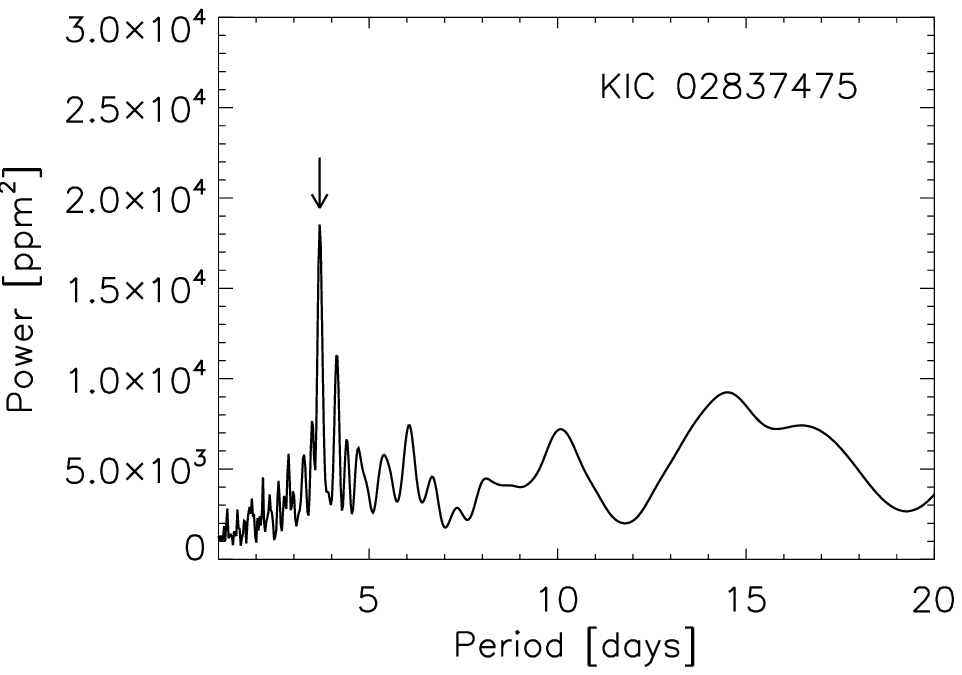}
\includegraphics[width=6.0cm]{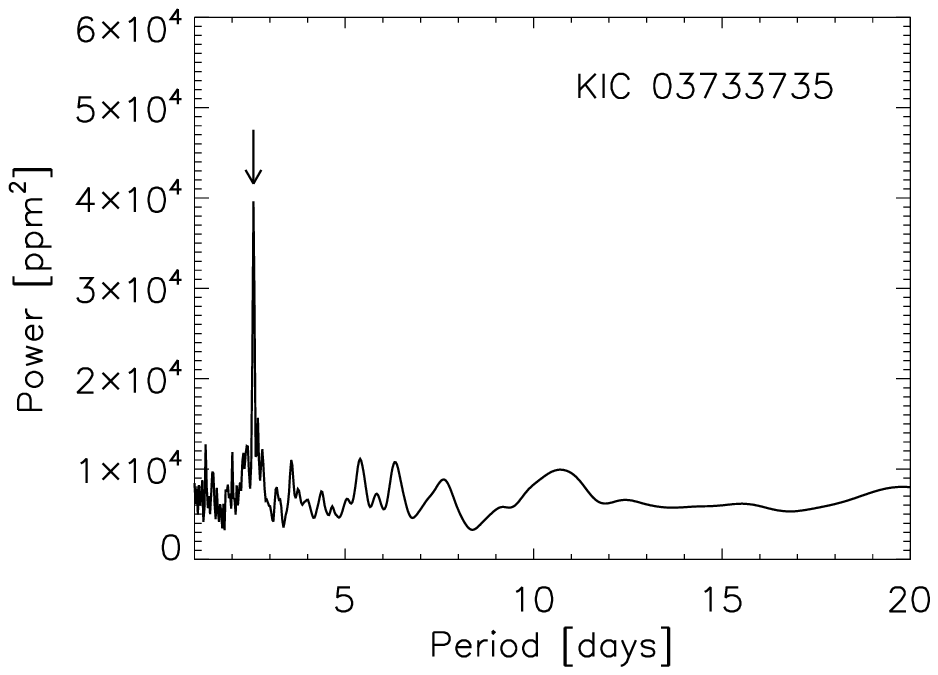}}}

\centerline{\hbox{\includegraphics[width=6.0cm]{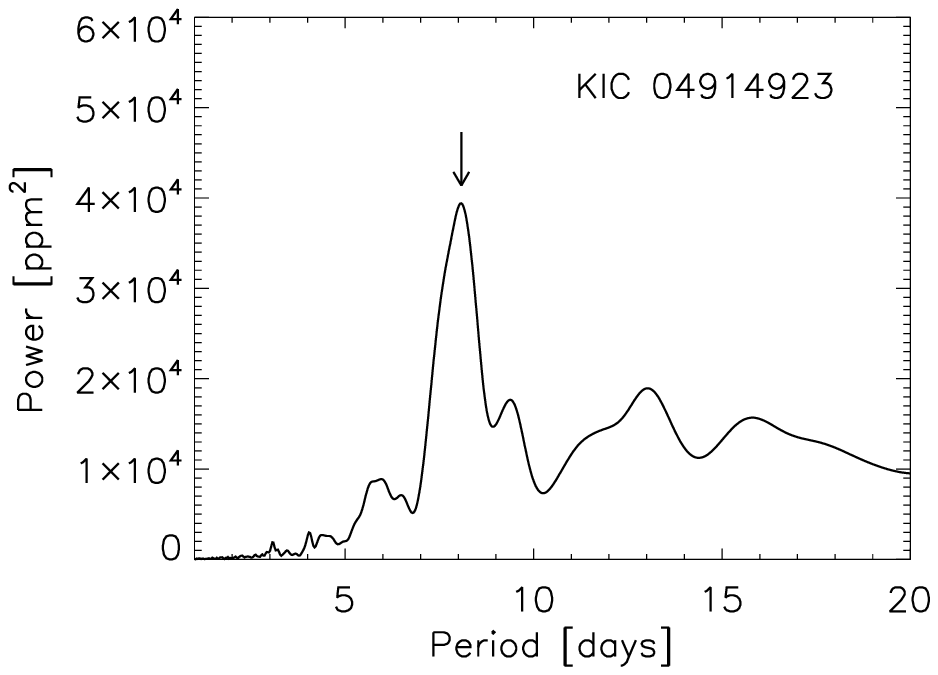}
\includegraphics[width=6.0cm]{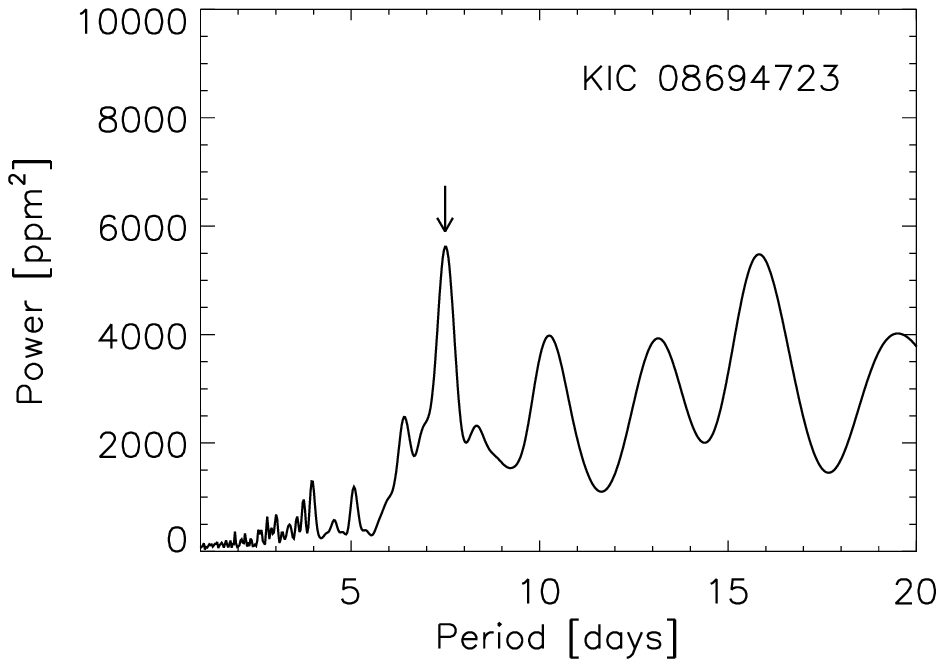}
\includegraphics[width=6.0cm]{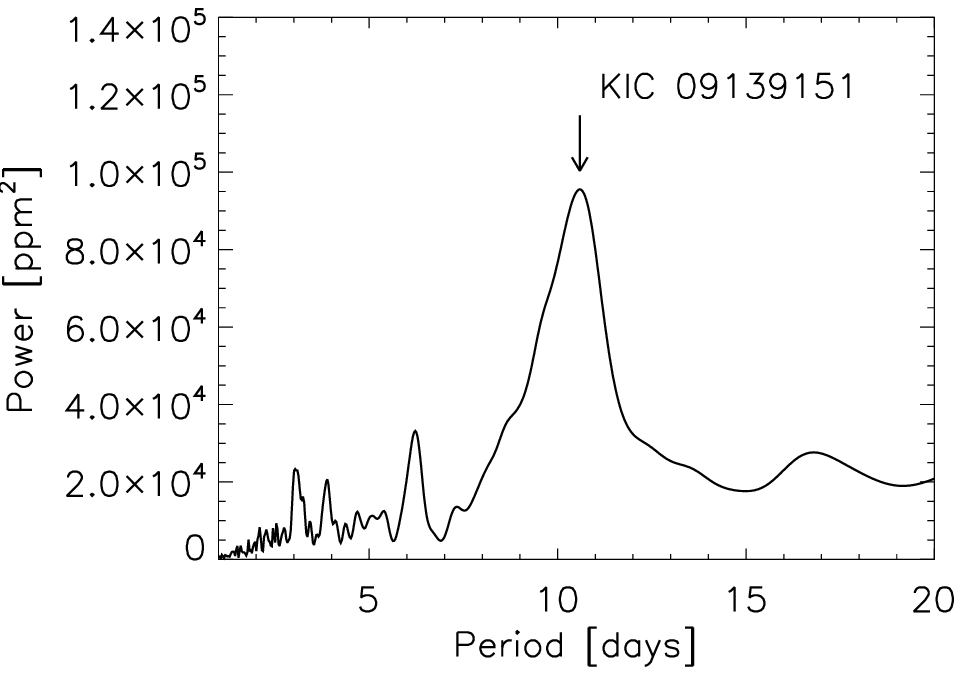}}}

\centerline{\hbox{\includegraphics[width=6.0cm]{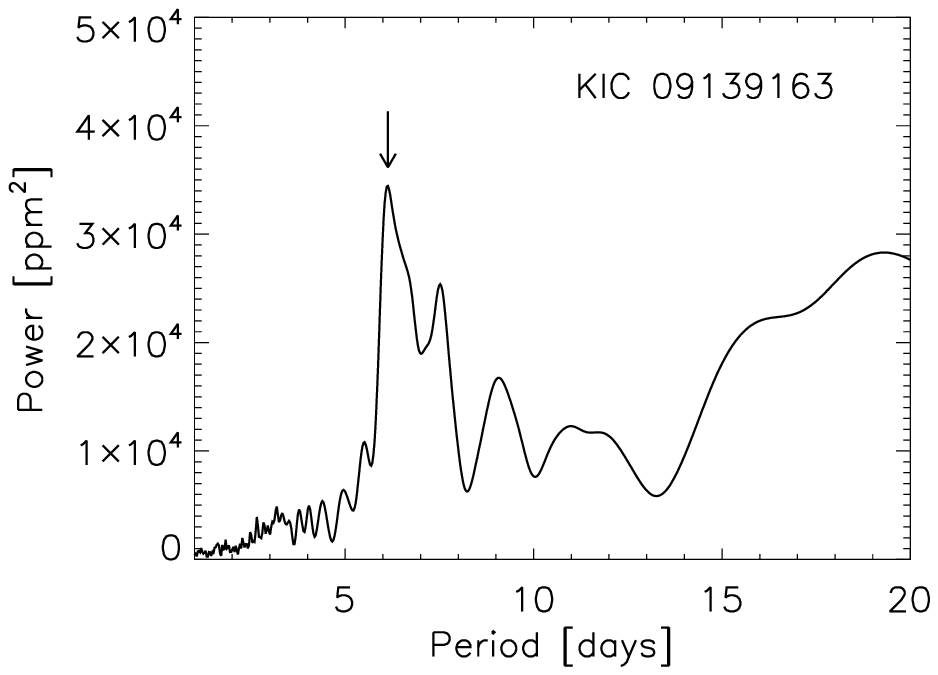} 
\includegraphics[width=6.0cm]{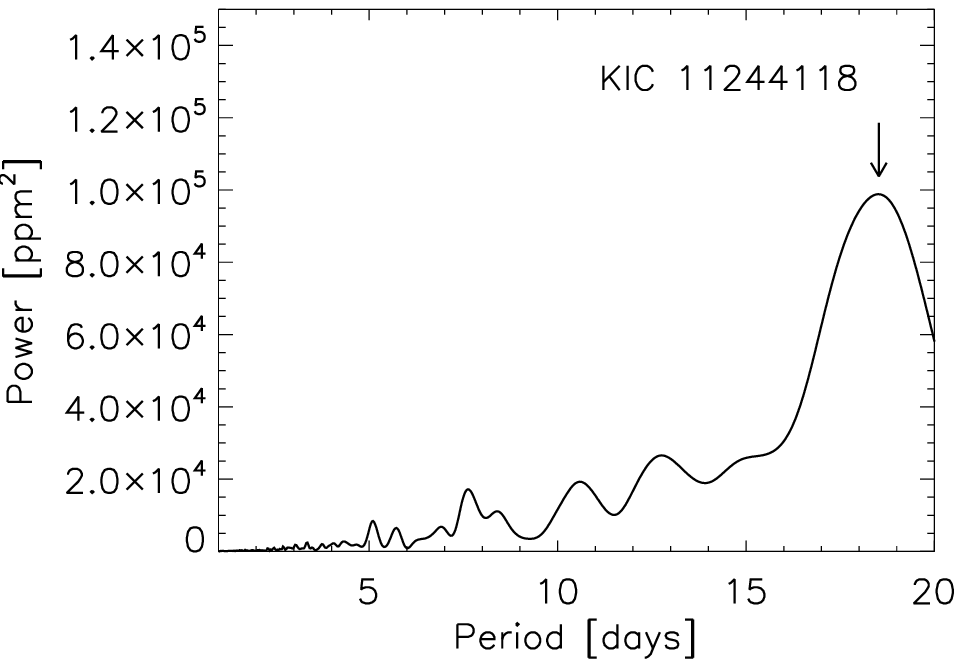}
\includegraphics[width=6.0cm]{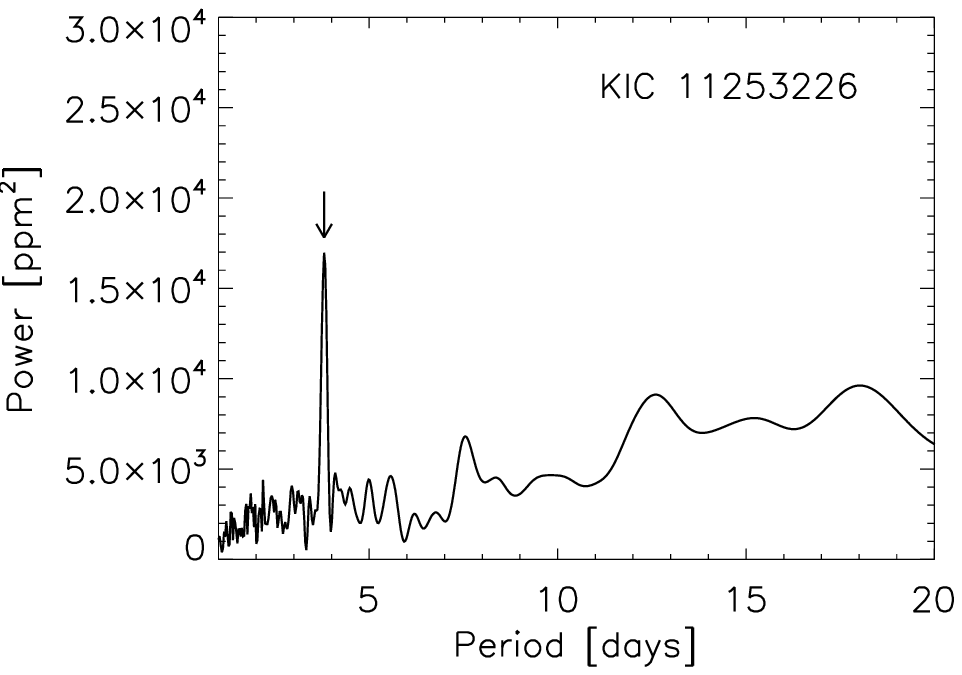}}}

\includegraphics[width=6.0cm]{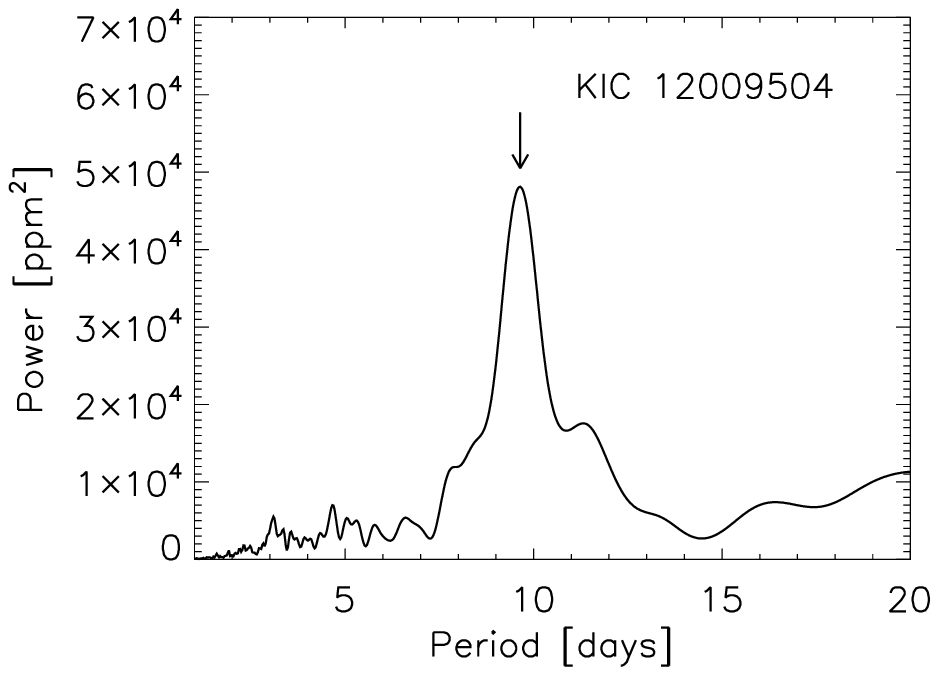}
\caption{Periodograms of the stars showing rotational modulation of their light curves. There arrows indicates the adopted rotation period. The adopted period is identical to the highest peak in all the periodograms.}                                   
\end{figure*} 
\end{center}

\label{lastpage}

\end{document}